\documentclass[twocolumn,aps,prb,superscriptaddress,floatfix]{revtex4-2}
\usepackage[T1]{fontenc}
\usepackage[utf8]{inputenc}
\usepackage[english]{babel}
\usepackage{color}
\usepackage{float}
\usepackage{braket}
\usepackage{amsmath}
\usepackage{amstext}
\usepackage{amssymb}
\usepackage{graphicx}
\usepackage{bbold}
\usepackage[unicode=true,pdfusetitle,
bookmarks=true,bookmarksnumbered=false,bookmarksopen=false,
breaklinks=false,pdfborder={0 0 1},backref=false,colorlinks=false]
{hyperref}
\newcommand{\angstrom}{\textup{\AA}}
\hypersetup{
	colorlinks,linkcolor=blue,citecolor=blue,urlcolor=blue}
\usepackage{soul}
\makeatletter
\date{}

\makeatother
\setul{0.5ex}{0.3ex}
\definecolor{Blue}{rgb}{0,0.0,1}
\setulcolor{Blue}
\usepackage{babel}
\newcommand{\new}[1]{{\color{blue}#1}}

\begin{document} 

\author{Tarik P. Cysne}
\affiliation{Instituto de F\'\i sica, Universidade Federal Fluminense, 24210-346 Niter\'oi RJ, Brazil} 
\email{tarik.cysne@gmail.com}

\author{Sayantika Bhowal}
\affiliation{Materials Theory, ETH Zurich, Wolfgang-Pauli-Strasse 27, 8093 Zurich, Switzerland}

\author{Giovanni Vignale}
\affiliation{Department of Physics and Astronomy, University of Missouri, Columbia, Missouri 65211, USA}

\author{Tatiana G. Rappoport}
\affiliation{Instituto de Telecomunicações, Instituto Superior Tecnico, University of Lisbon, Avenida Rovisco Pais 1, Lisboa, 1049001 Portugal}	
\affiliation{Instituto de F\'\i sica, Universidade Federal do Rio de Janeiro, C.P. 68528, 21941-972 Rio de Janeiro RJ, Brazil}

\title{Orbital Hall effect in bilayer transition metal dichalcogenides: From the intra-atomic approximation to the Bloch states orbital magnetic moment approach}

\begin{abstract}
Using an effective Dirac model, we study the orbital Hall effect (OHE) in bilayers of transition metal dichalcogenides with 2H stacking (2H-TMD). We use first-order perturbation theory in the interlayer coupling of the bilayer system to obtain analytical expressions for the orbital Hall conductivity in the linear response regime. We use two distinct descriptions of the orbital angular momentum (OAM) operator: The first one is the intra-atomic approximation that considers only the intrasite contribution to the OAM [Cysne et al. Phys. Rev. Lett. 126, 056601 (2021)]. The second one uses the Berry-phase formula of the orbital (valley) magnetic moment to describe the OAM operator [Bhowal and Vignale, Phys. Rev. B 103, 195309 (2021)]. This approach includes both intersite and intrasite contributions to the OAM. Our results suggest that the two approaches agree qualitatively in describing the OHE in bilayers of 2H-TMDs, although they present some quantitative differences. We also show that interlayer coupling plays an essential role in understanding the OHE in the unbiased bilayer of 2H-TMD. This coupling causes the Bloch states to become bonding (antibonding) combinations of states of individual layers, demanding the consideration of the non-Abelian structure of the orbital magnetic moment to the occurrence of OHE. As we discuss throughout the work, the emerging picture of transport of OAM in the unbiased bilayer of 2H-TMDs based on OHE is very different from the usual picture based on the valley Hall effect, shedding new lights on previous experimental results. We also discuss the effect of the inclusion of a gate-voltage bias in the bilayer system. Our work gives support to recent theoretical predictions on OHE in two-dimensional materials.
\end{abstract}
\maketitle


\section{Introduction} 

The orbital Hall effect (OHE) consists of the transverse flow of orbital angular momentum (OAM) as a response to the application of a longitudinal electric field. It is a phenomenon analogous to the spin Hall effect but, contrary to the latter, the OHE does not require the existence of strong spin-orbit coupling in the material. Despite being predicted more than a decade ago \cite{OHEBernevig}, only recently the OHE has gained significant attention by the condensed-matter community, which contrasts with the spin Hall effect that has been the focus of intense research in the last twenty years. The first theoretical studies on the OHE had focused on three-dimensional metallic systems \cite{OHEorigins-1, OHEorigins-2, OHEorigins-3}, where a mechanism based on the electrical response of orbital texture in the materials were introduced \cite{OrbitalTexture, OHEmetals, Negative-OHE}. Experiments on orbital-torque had found signatures of the existence of the OHE \cite{Exp_orbital_Torque-1, Exp_orbital_Torque-2, Exp_orbital_Torque-3, Orbital-torque-1, Orbital-torque-2}. In addition, a recent experiment reported the direct measurement of OHE \cite{Exp-OHE-1}, opening the way for the development of the field of orbitronics \cite{Go-Review, Orbital-Hall-Phonon, Orbital-Rashba, Han-Orbitaldynamics}.

The interest in physical phenomena related to OAM, particularly the OHE, has gained substantial push in the community of two-dimensional (2D) materials \cite{OHE_VHE_PRB_Imaging, Thermoelectric-Orbital-Mag, OrbitalMagnetization-Xiao, OpticallyControlOrbitronic, GraphaneOHE-2010, OrbitalPhotoCurrent, OrbitalPhotoCurrent-2, Us4-PRB, Orbital-Ordering2D-1, Orbital-Ordering2D-2, Orbital-Ordering2D-3, OrbitalMagnetoelectric-Bilayer-graphene}. Theoretical calculations predict the existence of the orbital textures that give origin to the OHE in many 2D multiorbital materials \cite{Luis-PRL, Us1-PRB, Us2-PRBR, BoropheneTexture-DFT}. In addition, recent experiments confirmed the existence of these textures for some of these 2D materials \cite{Orbital-Texture-Exp-TMD-1, Orbital-Texture-Exp-TMD-2, Orbital-Texture-Exp-TMD-3}. Among many 2D materials, the family of transition metal dichalcogenides (TMDs) has attracted prominent interest. When crystallized in H structural phase, the TMDs are semiconductors with a large gap. The OAM physics, in these materials, is well described by the electronic states near valleys of the Brillouin zone (BZ). Previous works have shown that TMDs exhibit a strong OHE \cite{OHE_Bhowal_1, OHE_Bhowal_2, OHE_VHE_PRB_Imaging}, even inside its insulating gap \cite{Us1-PRB, Us2-PRBR}, in which it is possible to attribute an orbital-Chern number \cite{Us3-PRL}. The zigzag nanoribbons of TMDs have orbitally-polarized edge-states that may transport the orbital Hall currents \cite{Edgestates-TMD-1, Edgestates-TMD-2}. Added to this, the OHE allows the flux of OAM in centrosymmetric bilayers of TMDs with 2H stacking (bilayer of 2H-TMDs) \cite{Us3-PRL}. Previous studies often neglected this fact, interpreting the transport of OAM in terms of the valley Hall effect \cite{VHE-Momentum_2, VHE-Momentum_3, VHE-Momentum_4, VHE-Momentum_5}.

Most of the literature on OHE uses of the atomic representation of the OAM operator, also known as intra-atomic approximation. Due to its low computational cost and easy implementation, this approximation is widely used to describe the orbital properties of materials. The intra-atomic approximation to OAM frequently gives satisfactory results but neglects contributions generated by the movement of electrons in the intersite region of the solid \cite{ModTheo-1, ModTheo-2, ModTheo-3, ModTheo-4, vanderbiltBook}. The relevance of intersite contributions and the validity of the intra-atomic approximation strongly depend on the specificities of the material \cite{Importanceof-ModTheo-1, Importanceof-ModTheo-2, Importanceof-ModTheo-3, Importanceof-ModTheo-4}. In systems formed by atoms with well-localized outer shells, the approximation can be conceptually justified \cite{vanderbiltBook}. Recently, Bhowal and Vignale \cite{OHE_Bhowal-Vignale} introduced a scheme to take into account, on equal footing, the intra-atomic (intrasite) and the correction due to the extended nature of electronic wave functions (intersite) using a description of the OAM operator based on the Berry-phase formula for orbital (valley) magnetic moment. They used the method to express the valley Hall effect on the gapped graphene model \cite{VHE-graphene-XIAO} in a more transparent picture of OHE. In this work, we use the method introduced in Ref. \cite{OHE_Bhowal-Vignale} to study the OHE in bilayers of 2H-TMDs and compare it with results obtained within intra-atomic approximation \cite{Us3-PRL}. As we show through the work, both approaches agree qualitatively, giving robustness to the recent predictions regarding the transport of OAM in 2D materials. For clarity and transparency, we use an effective Dirac model to describe the low-energy physics of bilayer of 2H-TMD \cite{Bukard-VHE-bilayer, Low-Energy-bilayer}, allowing us to obtain analytical expressions for orbital Hall conductivity in the two schemes. 

Our work contains two important physical messages for the field of the OHE in 2D materials. The first message is the qualitative agreement between intra-atomic approximation and the Bloch states orbital magnetic moment approach in predicting an orbital Hall insulating plateau in centrosymmetric bilayers of 2H-TMDs with the height expressed in Eqs. (\ref{SigmaAtomicGap}, \ref{SigmaBerryGap}). Despite the quantitative difference in the results from the two methods, the orbital Hall conductivity agrees qualitatively, as is shown in Figs. \ref{fig:fig2} and \ref{fig:fig3}. To obtain the finite orbital Hall plateau in Eq. (\ref{SigmaBerryGap}) becomes necessary to consider the non-Abelian nature of the orbital magnetic moment operator [see appendix \ref{Appendix}]. This necessity comes from finite interlayer hopping ($t_{\perp}$) in bilayer 2H-TMDs that connects the Hilbert spaces of each layer. The second important message of the work is the reinforcement of the OHE as a better description of the transport of OAM. The conventional view based on the valley Hall effect presents many conceptual problems which forbid the possibility of transport of OAM in centrosymmetric non-magnetic systems.

We organize the paper as follows: In sec \ref{sec2}, we describe the effective Dirac model used in this work to explore the OHE in the bilayer of 2H-TMDs. We also perform first-order perturbation theory in the interlayer coupling, obtaining the eigenvectors and energies of the bilayer with no applied gate voltage (unbiased bilayer). In Sec \ref{sec3}, we consider these perturbed eigenvectors and energies to study the OHE for unbiased bilayer of 2H-TMD in linear-response regime. We consider both the intra-atomic approximation \cite{Us3-PRL} and the Bloch state orbital magnetic moment description of OAM \cite{OHE_Bhowal-Vignale} and compare the two results. In Sec. \ref{sec4}, we introduce an asymmetry between layers of the bilayer 2H-TMD by adding a gate potential (bias) in the system and performing a similar analysis of Secs. \ref{sec2} and \ref{sec3}. In Sec \ref{sec5}, we present the final remarks and conclusions of the work. We also include an appendix \ref{Appendix} where we review the construction of the orbital magnetic moment matrix associated with Bloch states used in Ref. \cite{OHE_Bhowal-Vignale} to study the OHE in gapped graphene. In addition, in the appendix \ref{AppendixB}, we show some technical details on the computation of orbital-current in the two approaches.


\section{Effective model for unbiased bilayer TMDs: Perturbation theory \label{sec2}} 

\begin{figure}[h]
	\centering
	 \includegraphics[width=1.0\linewidth,clip]{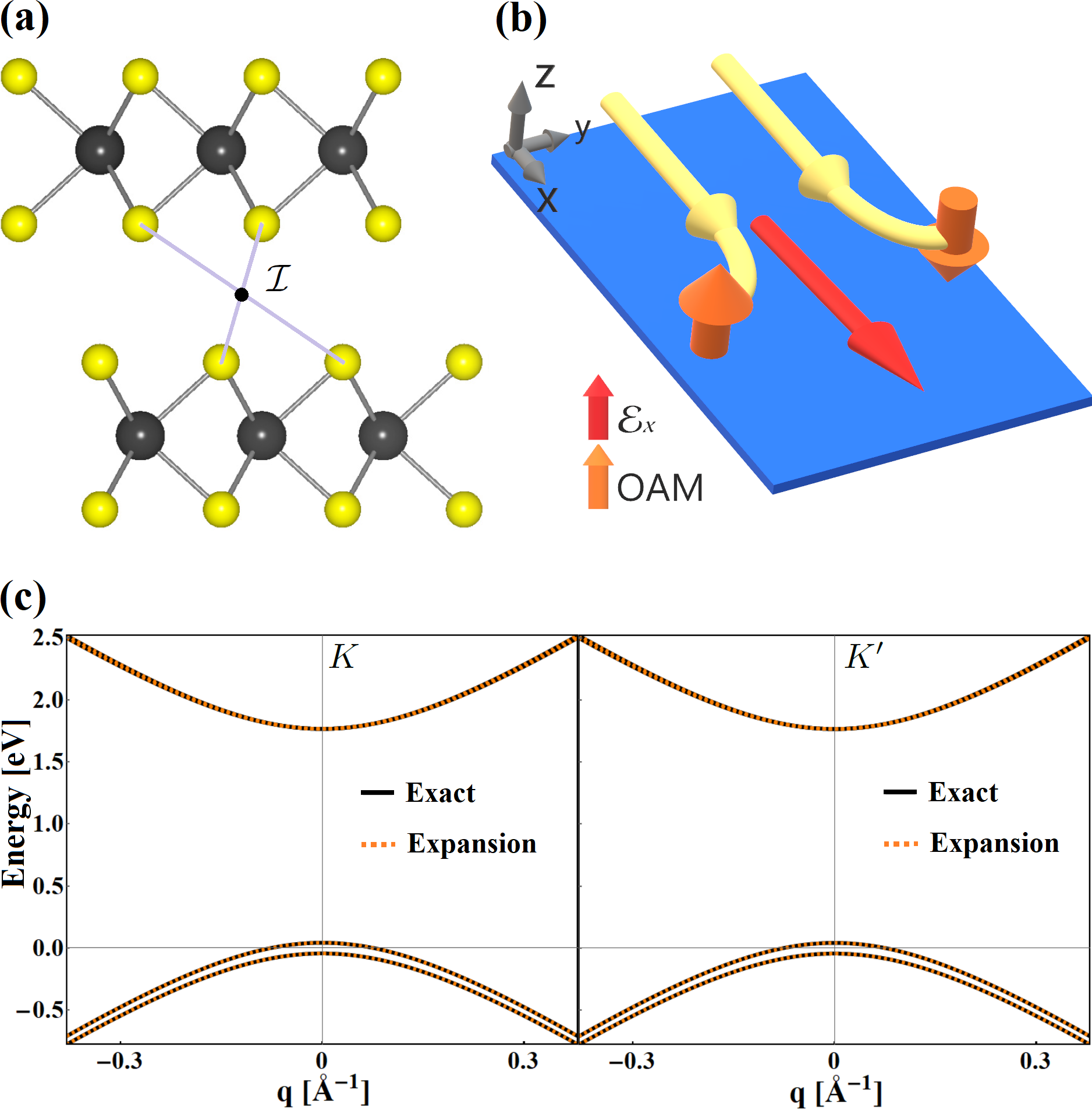}   
  
	\caption{(a) Side view of a bilayer of TMD with 2H stacking. Each monolayer of TMD in the H structural phase is composed of an atomic plane of transition metal (black spheres) sandwiched by two atomic planes of chalcogens (yellow spheres). The top and the bottom layers have a relative rotation angle of 180 degrees, making the bilayer system centrosymmetric. $\mathcal{I}$ represents a point of inversion symmetry of the bilayer. (b) Schematic representation of OHE, i.e., the transverse flux of orbital-current induced by a longitudinal ($\hat{x}$ direction) electric field. In the picture, the orbital-current flows in the $-\hat{y}$ direction and has OAM polarized in the $\hat{z}$ direction. (c) The energy spectrum of the bilayer of 2H-TMD near valleys $K$ (left) and $K'$ (right) of BZ. The solid black curve shows the spectrum obtained by numerical diagonalization of the Hamiltonian of Eq.(\ref{Heff}) without considering the spin-orbit coupling ($\lambda \rightarrow 0$). The dashed orange curve shows the energy spectra obtained by first-order perturbation theory expansion in interlayer hopping $t_{\perp}$ [Eqs. (\ref{CorrectedEnergies}, \ref{correctionVal}, \ref{correctionCond})].}
	\label{fig:fig1}
\end{figure}

The OAM physics of the TMDs is dominated by the points $K$ and $K'$ of BZ. At these points, the wave function at the top of the valence band and the bottom of the conduction band is formed by the orbitals $d_{z^2}$, $d_{x^2-y^2}$ and $d_{xy}$ of the transition metal atoms \cite{Xiao_PRL_2012, 3-Bands-Model, Low-Energy-bilayer, 3Bands-Mockli-Khodas, MarcosGuimaraes-TMD-Symmetries, Low-energy_x_Wannier-TMD}. We follow Refs. \cite{Bukard-VHE-bilayer, Low-Energy-bilayer} to build a simplified tight-binding (TB) model Hamiltonian in reciprocal space, which we expand up to first order in the electronic momentum around the valleys located at $\vec{K}=(4\pi/3a)\hat{x}$ and $\vec{K}'=-\vec{K}$. This procedure leads to the following Hamiltonian:
\begin{eqnarray}
H(\vec{q})=\begin{bmatrix}
\Delta & \gamma_+  & 0 & 0\\
  \gamma_- & -\tau s_z \lambda & 0 & t_{\perp}\\
0 & 0 & \Delta & \gamma_-\\
0 & t_{\perp} & \gamma_+ & \tau s_z \lambda
\end{bmatrix},
\label{Heff}
\end{eqnarray}
where  $\gamma_\pm = at(\tau q_x \pm i q_y)$, $\tau=\pm 1$ is the valley quantum number associated with valleys $K$ and $K'$, respectively. The TB basis of the Hamiltonian of Eq. (\ref{Heff}) is $\beta_{tb}=\{ \big|d^1_{z^2}\big>,\big( \big|d^1_{x^2-y^2}\big>-i\tau \big|d^1_{xy}\big> \big)/\sqrt{2}, \big|d^2_{z^2}\big>, \big( \big|d^2_{x^2-y^2}\big>+i\tau \big|d^2_{xy}\big> \big)/\sqrt{2}\}$, where the superscripts 1 and 2 specify the two layers of the bilayer, respectively [see Fig. \ref{fig:fig1} (a)]. Here, $\vec{k} = \vec{q} + \tau \vec{K}$ where $\vec{q}$ represents the wavevector relative to valleys and $s_z$ denotes the usual Pauli matrix associated with the spin degree of freedom. This model can be easily applied to describe bilayers of compounds of the TMD family in the trigonal prismatic phase (H), with 2H stacking, as represented in Fig. \ref{fig:fig1} (a). In this paper, we will consider the case of bilayers of MoS$_2$, an archetypical member of the family of TMDs. The parameters of effective Hamiltonian can be obtained by adjusting results given by density functional theory calculations. For 2H-MoS$_2$ bilayers, we obtain the band-gap $\Delta=1.766 \text{eV}$, the lattice constant $a=3.160 \angstrom$, the intralayer nearest-neighbor hopping $t=1.137$ eV, and the interlayer hopping $t_{\perp}=0.043$ eV \cite{Low-Energy-bilayer}. We also obtain a spin-orbit coupling $\lambda=0.073$ eV \cite{Low-Energy-bilayer}. Here we are interested in describing the OHE in bilayers of TMDs, that is not significantly affected by spin-orbit coupling. For this reason, we set $\lambda \rightarrow 0$ and introduced a spin-degeneracy factor $g_s=2$ in all results presented in this work. 
 
To perform the perturbation theory expansion, we separate the Hamiltonian into two terms, $H_0(\vec{q})$, of decoupled layers of bilayer 2H-TMD,
\begin{eqnarray}
H_0(\vec{q})&=&\begin{bmatrix}
\Delta & \gamma_+  & 0 & 0\\
  \gamma_- & 0 & 0 & 0\\
0 & 0 & \Delta & \gamma_-\\
0 & 0 & \gamma_+ & 0
\end{bmatrix},\label{H0}
\end{eqnarray}
and the interlayer coupling term ($H_1$), treated as a perturbation:
\begin{eqnarray}
H_1&=&\begin{bmatrix}
0 & 0  & 0 & 0\\
0 & 0 & 0 & t_{\perp}\\
0 & 0 & 0 & 0\\
0 & t_{\perp} & 0 & 0
\end{bmatrix}.
\label{H1}
\end{eqnarray}
It is straightforward to obtain the eigenvectors and energies of the unperturbed Hamiltonian $H_0$. The conduction and valence bands of $H_0$ are doubly degenerate, having the energy-dispersion
\begin{eqnarray}
\epsilon^0_{v(c)}(q)=\frac{1}{2}\left(\Delta \mp \sqrt{\Delta^2+4a^2t^2q^2} \right), \label{E0}
\end{eqnarray}
where $q=\sqrt{q_x^2+q_y^2}$. The unperturbed eigenvectors of the valence ($v$) band are
\begin{eqnarray}
\big|\psi_{1,\tau,v}\rangle= \mathcal{N}_v(q) \left(\frac{\epsilon^0_{v}(q)}{\gamma_-},1,0,0\right)^T, \label{PsiVal01} \\
\big|\psi_{2,\tau,v}\rangle= \mathcal{N}_v(q) \left(0,0, \frac{\epsilon^0_{v}(q)}{\gamma_+},1\right)^T, \label{PsiVal02} 
\end{eqnarray}
while the conduction ($c$) band eigenvectors are
\begin{eqnarray}
\big|\psi_{1,\tau,c}\rangle=\mathcal{N}_c(q) \left(\frac{\epsilon^0_c(q)}{\gamma_-},1,0,0\right)^T, \label{PsiCond01} \\
\big|\psi_{2,\tau,c}\rangle=\mathcal{N}_c(q) \left(0,0,\frac{\epsilon^0_c(q)}{\gamma_+},1\right)^T. \label{PsiCond02} 
\end{eqnarray}
The normalization factors are $\mathcal{N}_{v(c)}(q)=\left[1+(\epsilon^0_{v(c)}(q))^2/(a t q)^2 \right]^{-1/2}$ and the superscript $T$ means the application of transpose operation to obtain column vectors. Note that the wavefunction of states on Eqs. (\ref{PsiVal01}, \ref{PsiCond01}) are localized on layer 1 and the wavefunctions of states on Eqs. (\ref{PsiVal02}, \ref{PsiCond02}) are localized on layer 2. The factors $\gamma_{\pm}$ defined above contain the dependence on valley quantum number $\tau$. To include the effect of interlayer hopping $t_{\perp}$, we apply standard degenerate perturbation theory by constructing the matrix $\langle \chi \big|H_1\big| \phi \rangle$ with states of valence [Eqs. (\ref{PsiVal01}, \ref{PsiVal02})] and conduction [Eqs. (\ref{PsiCond01}, \ref{PsiCond02})] bands subspace and calculating its eigenvectors to obtain a linear combination of these states. With this procedure, the effect of interlayer coupling translates into the formation of bonding ($+$) and antibonding ($-$) linear combinations of the eigenstates of individual layers on valence and conduction bands subspace:
\begin{eqnarray}
\big|\Phi_{\pm,\tau,v(c)}\rangle =\frac{1}{\sqrt{2}} \left(\big|\psi_{1,\tau,v(c)}\rangle \pm \big|\psi_{2,\tau,v(c)}\rangle \right). \label{CorrectedStates}
\end{eqnarray}
With these states, we compute the first-order correction of interlayer hopping $t_{\perp}$ to energies $\delta\epsilon_{v(c),\pm}(q) = \langle \Phi_{\pm,\tau,v(c)}\big| H_1\big|\Phi_{\pm,\tau,v(c)}\rangle$, obtaining
\begin{eqnarray}
\bar{\epsilon}_{v(c),\pm}(q)=\epsilon^0_{v(c)}(q)+\delta\epsilon_{v(c),\pm}(q), \label{CorrectedEnergies}
\end{eqnarray}
where 
\begin{eqnarray}
\delta\epsilon_{v,\pm}(q)=\pm \frac{t_{\perp}}{2} \left(1+\frac{\Delta}{\sqrt{\Delta^2+4a^2t^2q^2}}\right), \label{correctionVal} \\
\delta\epsilon_{c,\pm}(q)=\pm \frac{t_{\perp}}{2} \left(1-\frac{\Delta}{\sqrt{\Delta^2+4a^2t^2q^2}}\right). \label{correctionCond}
\end{eqnarray}
Note that energies of Eqs. (\ref{CorrectedEnergies}, \ref{correctionVal}, \ref{correctionCond}) are the same for different valleys due to time-reversal symmetry and the absence of spin-orbit coupling. In Fig. \ref{fig:fig1} (c), we show the comparison between the exact energy spectrum obtained by numerical diagonalization of the Hamiltonian of Eq. (\ref{Heff}) with $\lambda\rightarrow 0$ and the one obtained via perturbation theory [Eqs. (\ref{CorrectedEnergies}, \ref{correctionVal}, \ref{correctionCond})]. They agree very well within the regime of small wavevectors. It is interesting to note from Fig. \ref{fig:fig1} and Eqs. (\ref{correctionVal}, \ref{correctionCond}) that at the Dirac points ($q\rightarrow 0$), the energy-splitting induced by interlayer hopping $t_{\perp}$ occurs at the valence band but not on the conduction band. In realistic bilayers, the finite spin-orbit coupling causes a small energy-splitting in the conduction band at the Dirac point \cite{Bukard-VHE-bilayer, Low-Energy-bilayer}. This energy-splitting of the conduction band does not occur in the spectra obtained with $\lambda \rightarrow 0$ in Fig. \ref{fig:fig1} (c). As we mentioned before, the spin-orbit interaction is not relevant to the description of the OAM transport in this system being neglected in this work.

\section{Orbital Hall effect of unbiased bilayer TMDs \label{sec3}}

\subsection{Linear response theory for orbital Hall current}
To study the OHE, we use the formalism of linear-response theory where the orbital Hall current, that flow in the y-direction with OAM polarized in the z-direction (out-of-plane), generated by a longitudinal (x-direction) electric field is proportional to the orbital Hall conductivity (OHC), $\mathcal{J}^{X_z}_y=\sigma^{X_z}_{OH}\mathcal{E}_x$ [see Fig. \ref{fig:fig1} (b)], where OHC is given by \cite{OHEBernevig,OHEorigins-1, OHEorigins-2, OHEorigins-3, OpticallyControlOrbitronic, GraphaneOHE-2010, OrbitalPhotoCurrent, OrbitalPhotoCurrent-2},
\begin{eqnarray}
\sigma^{X_z}_{OH}=e\sum_n\int \frac{d^2k}{(2\pi)^2} f_{n,\vec{k}} \Omega^{X_z}_{n,\vec{k}}, \label{SigmaLR}
\end{eqnarray}
where $f_{n,\vec{k}}=\Theta (E_f-E_{n,\vec{k}})$ is the Fermi-Dirac distribution at zero temperature and Fermi energy $E_f$, and the orbital-weighted Berry curvature is given by
\begin{eqnarray}
\frac{\Omega^{X_z}_{n,\vec{k}}}{2\hbar}=\sum_{m\neq n} \text{Im} \left[ \frac{\langle u_{n,\vec{k}}\big|\hat{v}_x(\vec{k})\big| u_{m,\vec{k}}\rangle \langle u_{m,\vec{k}}\big|\hat{J}^{X_z}_y(\vec{k})\big| u_{n,\vec{k}}\rangle}{\left(E_{n,\vec{k}}-E_{m,\vec{k}}\right)^2} \right] . \nonumber \\
\label{OBC}
\end{eqnarray}
In the above equations, $E_{n(m),\vec{k}}$ are the energies of the eigenstates $\big|u_{n(m),\vec{k}}\rangle$ of the electronic Hamiltonian evaluated in wavevector space $\hat{\mathcal{H}}(\vec{k})$. The velocity operator in the x(y)-direction is defined by $\hat{v}_{x(y)}(\vec{k})=\hbar^{-1}\partial \hat{\mathcal{H}}(\vec{k})/\partial k_{x(y)}$. We follow Refs. \cite{Go-Review, OHEorigins-2} and define the OAM current operator $\hat{J}^{X_z}_y(\vec{k})= \left(X_z\hat{v}_{y}(\vec{k})+\hat{v}_{y}(\vec{k}) X_z\right)/2$. Here we apply this formalism to study the OHE in the bilayer TMD. As we mentioned in the introduction, we use two distinct descriptions of the OAM operator $X_z$ in this work. The first description, called intra-atomic approximation ($X_z=\hat{L}^{\text{Intra}}_z$), treats the OAM of electrons in a solid as an extension of atomic OAM. This approximation is widely used, especially in studies of OHE, due to its simple implementation \cite{vanderbiltBook}, often giving satisfactory results. The second description, recently introduced in the context of the OHE \cite{OHE_Bhowal-Vignale}, treats the OAM of electrons in the framework of the Berry phase formula of intrinsic orbital magnetic moment of Bloch states ($X_z=\hat{L}^{\text{Tot}}_z$). While the intra-atomic approximation takes into account only the intrasite contribution of electronic wave functions, the approach based on orbital magnetic moment includes both intrasite and intersite contributions to OAM \cite{Murakami-IntersiteIntrasite, ModTheo-5}. This justifies the use of superscript ``Tot'' used here in quantities described in the Bloch state orbital magnetic moment approach. We briefly mention that the orbital magnetic moment is also known as valley magnetic moment in the literature of 2D materials. Here, we adopt orbital magnetic moment following the nomenclature of Ref. \cite{OHE_Bhowal-Vignale}. In the remainder of this section, we compute the OHC for the bilayer TMD using the two descriptions of the OAM operator and compare the results. 

\subsection{OHE in the intra-atomic approximation \label{SecOHEAtomic}}
\begin{figure}[h]
	\centering
	 \includegraphics[width=0.9\linewidth,clip]{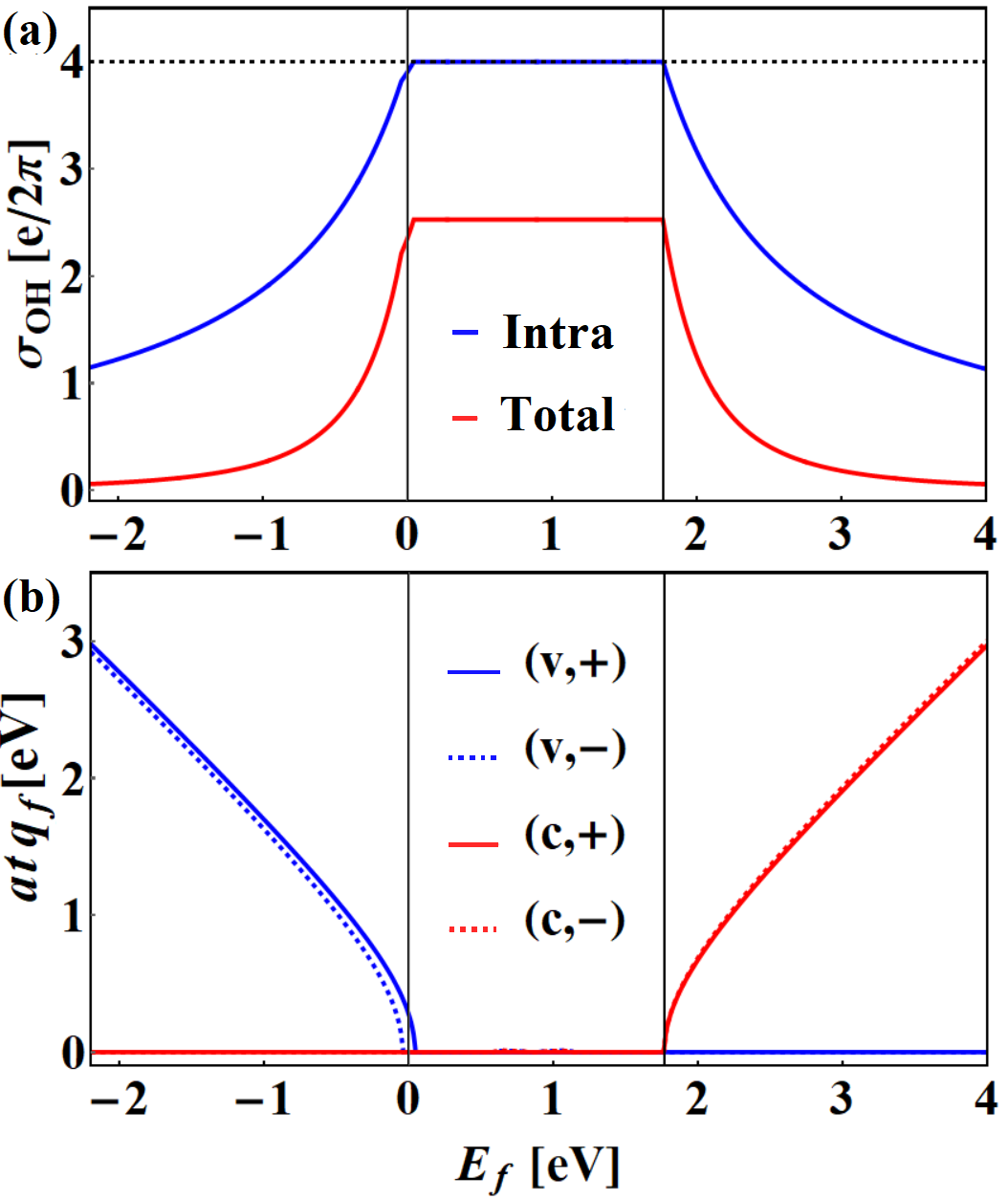}     
	\caption{(a) OHC as a function of the Fermi energy for unbiased bilayer of 2H-MoS$_2$. The two curves show the orbital conductivities calculated using intra-atomic approximation [Eqs. (\ref{SigmaAtom}), (\ref{SigmaAtomVal}), and (\ref{SigmaAtomCond})] and Bloch state orbital magnetic moment approach [Eqs. (\ref{SigmaBerry}), (\ref{SigmaBerryVal}), and (\ref{SigmaBerryCond})] for the OAM operator. (b) Fermi-momentum [Eq. (\ref{FermiMomenta})], as a function of Fermi-energy, for the valence and conduction band states in the unbiased bilayer of 2H-MoS$_2$.  The vertical continuous black lines in both panels delimit the insulating gap of the unbiased bilayer 2H-MoS$_2$ [Fig. \ref{fig:fig1} (c)]. The horizontal dashed black line in panel (a) signals the quantized orbital Hall conductivity in the intra-atomic approximation when Fermi-energy lies in the insulating gap [$\overline{\sigma}^{\text{Intra}}_{OH}=4  (e/2\pi)$, Eq. (\ref{SigmaAtomicGap})]. In panel (b), $a=3.160 \angstrom$ and $t=1.137 \text{eV}$ [see Sec. \ref{sec2}].}
	\label{fig:fig2}
\end{figure}

In the tight-binding basis $\beta_{tb}$ of the Hamiltonian of Eq. (\ref{Heff}), defined in sec. \ref{sec2}, the OAM operator in intra-atomic approximation for bilayer TMDs assumes the form \begin{eqnarray}
\hat{L}^{\text{Intra}}_z=\left( 2\hbar \tau \right)  \text{diag} \left(0, -1, 0, 1 \right). \label{LZAtomic}
\end{eqnarray}
It is straightforward to compute the velocity operators $\hat{v}_{x(y)}(\vec{q})$ from Eq. (\ref{Heff}). Using $v_y(\vec{q})$ and Eq. (\ref{LZAtomic}) to compute the operator $J^{z,\text{Intra}}_y$, we obtain the orbital-current operator in TB basis $\beta_{tb}$:
\begin{eqnarray}
J^{z,\text{Intra}}_y= at \tau\begin{bmatrix}
0 & -i  & 0 & 0\\
i  & 0 & 0 & 0\\
0 & 0 & 0 & -i \\
0 & 0 & i  & 0
\end{bmatrix}. \label{JorbAtomic}
\end{eqnarray}
We use this operator together with the corrected energies [Eqs. (\ref{CorrectedEnergies}, \ref{correctionVal}, \ref{correctionCond})] and states [Eq. (\ref{CorrectedStates})] given by the first-order perturbation theory to compute the orbital-weighted Berry-curvature in the intra-atomic approximation $\Omega^{\text{Intra}}_{c(v),\pm}(\vec{q})$. The analytical expressions for these curvatures are slightly cumbersome but they assume a simple form after performing the integral over azimuthal angle $\theta$ (where, $q_x= q\cos (\theta)$ and, $q_y= q\sin (\theta)$) and expanding in linear order on $t_{\perp}/\Delta$. This procedure leads us to the integral
\begin{eqnarray}
\int_0^{2\pi}&& \frac{d\theta}{(2\pi)} \Omega^{\text{Intra}}_{v,\pm}(\vec{q}) \approx -\frac{2a^2t^2\Delta}{\left(4a^2t^2q^2+\Delta^2\right)^{3/2}} \nonumber \\
&&\mp t_{\perp}\frac{2a^2t^2\Delta\left(4a^2t^2q^2+\Delta^2+\Delta\sqrt{4a^2t^2q^2 +\Delta^2} \right)}{\left( 4a^2t^2q^2+\Delta^2\right)^3} \nonumber \\ \label{AtomicOmegaVal}
\end{eqnarray}
for the valence band and 
\begin{eqnarray}
\int_0^{2\pi}&& \frac{d\theta}{(2\pi)} \Omega^{\text{Intra}}_{c,\pm}(\vec{q}) \approx \frac{2a^2t^2\Delta}{\left(4a^2t^2q^2+\Delta^2\right)^{3/2}} \nonumber \\
&&\mp t_{\perp}\frac{2a^2t^2\Delta\left(4a^2t^2q^2+\Delta^2-\Delta\sqrt{4a^2t^2q^2 +\Delta^2} \right)}{\left( 4a^2t^2q^2+\Delta^2\right)^3} \nonumber \\ \label{AtomicOmegaCond}
\end{eqnarray}
for the conduction band. Note that the right-hand side of these equations does not depend on the valley quantum number $\tau$. As a result, these curvatures add up when summed over the valleys, resulting in a finite OHC inside the insulating gap of the bilayer TMD \cite{Us1-PRB, Us2-PRBR,Us3-PRL}. We substitute Eqs. (\ref{AtomicOmegaVal}, \ref{AtomicOmegaCond}) in Eq. (\ref{SigmaLR}) and evaluate the radial integral, assuming the extrapolation $q\rightarrow \infty$ in the lower limit of integral. This corresponds to extrapolating the model of Eq. (\ref{Heff}) to small wavelengths. The highly peaked profile of the orbital-weighted Berry-curvature near the valleys of the BZ gives ground for this assumption. As a result we obtain the OHC in intra-atomic approximation:
\begin{eqnarray}
\sigma^{\text{Intra}}_{OH}(E_f)=\sum_{\nu=\pm} \left[ \sigma^{\text{Intra}}_{v,\nu} (E_f)+\sigma^{\text{Intra}}_{c,\nu} (E_f) \right], \label{SigmaAtom}
\end{eqnarray}
where the valence band contribution $\sigma^{\text{Intra}}_{v,\pm} (E_f)$ is given by
\begin{eqnarray}
\sigma^{\text{Intra}}_{v,\pm} (E_f)&&=g_s\frac{e}{2\pi} \Bigg[\frac{\Delta}{\sqrt{4a^2t^2q^2_{v,\pm}+\Delta^2}} \nonumber \\
&& \pm t_{\perp} \frac{\Delta \left(2\Delta+3\sqrt{4a^2t^2q^2_{v,\pm}+\Delta^2} \right)}{6\left(4a^2t^2q^2_{v,\pm}+\Delta^2 \right)^{3/2}}\Bigg], \label{SigmaAtomVal}
\end{eqnarray} 
and, the contribution of conduction band $\sigma^{\text{Intra}}_{c,\pm} (E_f)$ is 
\begin{eqnarray}
&&\sigma^{\text{Intra}}_{c,\pm} (E_f)=-g_s \frac{e}{2\pi} \Bigg[1-\frac{\Delta}{\sqrt{4a^2t^2q^2_{c,\pm}+\Delta^2}} \nonumber \\
&&\mp t_{\perp}\frac{\left( \Delta^3+\left(2a^2t^2q^2_{c,\pm}-\Delta^2 \right)\sqrt{4a^2t^2q^2_{c,\pm}+\Delta^2} \right)}{3\Delta\left( 4a^2t^2q^2_{c,\pm}+\Delta^2\right)^{3/2}} \Bigg]. \label{SigmaAtomCond}
\end{eqnarray}
Note that we reinserted the spin-degeneracy factor $g_s=2$ in these equations. $q_{v(c),\pm}$ are Fermi moments of the bonding (+) and antibonding (-) states for valence (conduction) band. To compute the Fermi moments as a function of Fermi energy we invert Eq. (\ref{CorrectedEnergies}) and solve $\bar{\epsilon}_{v(c),\pm}(q_{v(c),\pm})=E_f$. We then obtain
\begin{eqnarray}
q_{v(c),\pm}(E_f)=\text{Re}\left[ \frac{\sqrt{\alpha_{\pm}(E_f)+\beta_{v(c),\pm}(E_f)\sqrt{\delta_{\pm}(E_f)}}}{at2\sqrt{2}} \right],\nonumber \\ \label{FermiMomenta}
\end{eqnarray}
where, $\alpha_{\pm}(E_f)=\left(4E_f^2+t^2_{\perp}-\Delta^2-4E_f\Delta\right)\pm 4t_{\perp}\left(\Delta-E_f\right)$, $\delta_{\pm}(E_f)=\left(4E_f^2+t_{\perp}^2-4E_f\Delta+\Delta^2\right)\pm t_{\perp} \left(6\Delta-4E_f\right)$, and $\beta_{v,\pm}=\left(\Delta \pm t_{\perp}-2E_f\right)$, $\beta_{c,\pm}=-\left(\Delta \pm t_{\perp}-2E_f\right)$. Fig. \ref{fig:fig2} (b)  shows the Fermi moments of Eq. (\ref{FermiMomenta}) as a function of Fermi energy for the typical parameters of bilayer MoS$_2$ introduced in Sec. \ref{sec2}.  

In Fig. \ref{fig:fig2} (a), we also show the orbital Hall conductivities given by Equations (\ref{SigmaAtom}), (\ref{SigmaAtomVal}), and (\ref{SigmaAtomCond}) for the unbiased bilayer MoS$_2$. The most striking feature of this curve is the quantization of the OHC in intra-atomic approximation when the Fermi-energy lies in the insulating gap:
\begin{eqnarray}
\overline{\sigma}^{\text{Intra}}_{OH}=4  \left(\frac{e}{2\pi}\right)= 2\mathcal{C}^{2l}_L  \left(\frac{e}{2\pi}\right).
\label{SigmaAtomicGap} 
\end{eqnarray}
This quantization occurs in first-order perturbation theory in interlayer hopping $t_{\perp}$. If one includes higher orders of interlayer hopping in the perturbative expansion, the height of the orbital Hall insulating plateau deviates slightly from this quantized value \cite{Us3-PRL}. A modification in this quantized value may also occur by including high-order terms in $\vec{q}$, such as trigonal warping \cite{Trigonal-Warping-1, Trigonal-Warping-2}, in the effective Dirac theory of Eq.(\ref{Heff}). As denoted by the second equality of Eq.(\ref{SigmaAtomicGap}), we can write the height of the quantized orbital Hall insulating plateau obtained here as $2\times \mathcal{C}^{2l}_L$, where factor 2 comes from the $d$-shell character of transition metals [Eq. (\ref{LZAtomic})], and $\mathcal{C}^{2l}_L$ is the orbital Chern-number introduced in Ref. \cite{Us3-PRL}. We can define this orbital Chern-number even in nonperturbative calculations \cite{Us3-PRL} using the method formalized in Refs. \cite{Spin-Chern-Number-1, Spin-Chern-Number-2, Spin-Chern-Number-3}. For the present case of the bilayer of 2H-TMD, the orbital Chern number is $\mathcal{C}^{2l}_L=2$. The same calculation presented above could be performed for the monolayer of TMD in the H structural phase (monolayer of H-TMD), giving an orbital Chern number $\mathcal{C}^{1l}_L=1$. The result of Eq.  (\ref{SigmaAtomicGap}) unveils a topological nature in the insulating gap of H-TMDs, despite being trivial in $\mathbb{Z}_2$-invariant. Zigzag nanoribbons of H-TMDs have orbitally-polarized edge states that cross the bulk gap \cite{Edgestates-TMD-1, Edgestates-TMD-2} and may be responsible for the transport of orbital Hall current.

\subsection{OHE in the Bloch state orbital magnetic moment description \label{SecOHEBerry}}
To contrast with the results of the previous subsection, we evaluate the OHC with the OAM described by the Berry phase formula of the Bloch state orbital magnetic moment. It was first shown by W. Kohn that Bloch electrons possess an intrinsic magnetic moment \cite{Bloch-Orbital-Moment-Khon-1}. Later, this intrinsic magnetic moment was connected with Berry phase theory and interpreted in a more transparent picture of the self-rotation of semiclassical wave-packet \cite{Bloch-Orbital-Moment-Chang-2}. The description of OAM operator in terms of the orbital magnetic moment of Bloch states was extensively explored \cite{Bloch-Orbital-Moment-Culcer-3, Bloch-Orbital-Moment-Review-1, Bloch-Orbital-Moment-Review-2, Thermoelectric-Orbital-Mag, OrbitalMagnetization-Xiao, OrbitalMagneticMoment-App-1, OrbitalMagneticMoment-App-2, WPacket-Culcer-2022}, also appearing in the context of the modern theory of orbital magnetization \cite{ModTheo-1, ModTheo-2, ModTheo-3, ModTheo-4, vanderbiltBook}. Recently, it was proposed in Ref. \cite{OHE_Bhowal-Vignale} that the long-debated valley Hall effect \cite{Roche-Valley-Hall} could be viewed as an OHE, using the description of the OAM operator in terms of the orbital magnetic moment. Here, we use the scheme introduced in this reference to study the OHE in bilayer TMDs. In appendix \ref{Appendix}, we review the general theory of orbital magnetic moment of Bloch states based on Refs. \cite{Bloch-Orbital-Moment-Khon-1, Bloch-Orbital-Moment-Chang-2, Bloch-Orbital-Moment-Culcer-3, Bloch-Orbital-Moment-Review-1, Bloch-Orbital-Moment-Review-2} and also obtain the operator used in this subsection. As detailed in the appendix \ref{Appsec2}, the correct description of the orbital magnetic moment in the unbiased bilayer system must consider its non-Abelian (matricial) nature.  On a TB basis $\beta_{tb}$, the orbital magnetic moment operator for the unbiased bilayer of 2H-TMD assumes the form of a diagonal matrix: \begin{eqnarray}
\hat{m}_{tb}(q)=\tau m_0(q)  \text{diag}\left(1, 1, -1, -1\right), \label{mtb}
\end{eqnarray}
where
\begin{eqnarray}
m_0(q)=\left( \frac{e}{\hbar}\right) \frac{a^2 t^2\Delta }{4a^2t^2q^2+\Delta^2}=\mu^{*}_B f\left( \frac{2atq}{\Delta} \right) \label{m0}
\end{eqnarray}
is the orbital magnetic moment of a massive Dirac fermion \cite{VHE-graphene-XIAO}. In the second equality of Eq. (\ref{m0}), we defined the dimensionless function $f(x)=(1+x^2)^{-1}$ and the renormalized Bohr magneton $\mu^*_B=(e\hbar) /(2m_e^*)$ with the effective mass $m_e^*=(\hbar^2 \Delta)/ (2a^2t^2)$. At the Dirac point $K$, $m_0$ reduces to the renormalized Bohr magneton, so that $m_0(q=0)=\mu^{*}_B$.

In the case of the orbital magnetic moment description, to calculate the OAM current operator, we need to multiply the magnetic moment by a constant to convert the magnetic moment to units of angular momentum. This constant depends on the Landé g-factor $g_L$, which can give rise to quantitative ambiguities in the case of materials with strong spin-orbit coupling. On the other hand, experiments observe magnetic moment accumulation.  So, in principle, $g_L$ is not necessary for the prediction of magnetic moment accumulation or non-local resistances. To avoid introducing this constant, one could alternatively describe the OHE in terms of a current of orbital magnetic moments $J^{z,\text{m}}_y(q)$. We can use the definition of the current operator in Eqs. (\ref{SigmaLR}, \ref{OBC}) where $X_z$ is the orbital magnetic moment operator. In this case the current follows directly from the orbital magnetic moment matrix $X_z=\hat{m}_{tb}(q)$ [Eq. (\ref{mtb})], 
\begin{eqnarray}
J^{z,\text{m}}_y(q)= -\frac{\tau at m_0(q)}{\hbar}\begin{bmatrix}
0 & -i  & 0 & 0\\
i  & 0 & 0 & 0\\
0 & 0 & 0 & -i \\
0 & 0 & i  & 0
\end{bmatrix}. \label{JOMMBerry}
\end{eqnarray}
In the case of the intra-atomic approximation, there is a trivial conversion of the OAM operator of Eq. (\ref{LZAtomic}) to the units of the orbital magnetic moment: $\hat{m}^{\text{Intra}}_z= \left(- \mu_B/\hbar \right) \hat{L}_z^{\text{Intra}}$. Nevertheless, both descriptions of the OHE using orbital magnetic moment current or OAM current have the same physical information in Eqs. (\ref{LZAtomic}, \ref{mtb}). 
 
Here, we follow the literature of OHE to connect with previous results and consider the OAM current. To convert the orbital magnetic moment operator of equation \ref{mtb} to units of angular momentum, we multiply it by the constant \cite{OHE_Bhowal-Vignale} $C_{\text{am}}=-\hbar g_L^{-1}\mu_B^{-1}=-2m_eg^{-1}_Le^{-1}$, where $g_L=1$, $\mu_B= (e\hbar)/(2m_e)$ is the atomic Bohr 
magneton and, $m_e$ is the electron rest mass. In TB basis $\beta_{tb}$, the OAM operator reads
\begin{eqnarray}
\hat{L}^{\text{Tot}}_z(q)= \left( \frac{\tau\hbar  m_0(q)}{\mu_B} \right)  \text{diag} \left(-1, -1, 1, 1 \right), \label{LZBerry}
\end{eqnarray}
that can be used to obtain the OAM current operator flowing in the y-direction
\begin{eqnarray}
J^{z,\text{Tot}}_y(q)= \frac{\tau atm_0(q)}{\mu_B}\begin{bmatrix}
0 & -i  & 0 & 0\\
i  & 0 & 0 & 0\\
0 & 0 & 0 & -i \\
0 & 0 & i  & 0
\end{bmatrix}. \label{JorbBerry}
\end{eqnarray}
At the Dirac point the OAM operator can be written in function of the ratio between the two Bohr magnetons: $\hat{L}^{\text{Tot}}_z (K_\tau)=\hbar \tau (\mu^{*}_B/\mu_B)\text{diag} (-1, -1, 1, 1)$. It is worth mentioning that the orbital-current operator within the Bloch state orbital magnetic moment approach [Eq. (\ref{JorbBerry})] has the same matrix structure as the operator obtained in the intra-atomic approximation [Eq. (\ref{JorbAtomic})]. This occurs despite the difference in the matrix structure of the OAM operators in both approaches [Eq. (\ref{LZAtomic}, \ref{LZBerry})]. It is easy to see how this occurs by representing the OAM operators in terms of a tensorial product of Pauli matrices related to orbital ($\sigma^i$) and layer ($\Sigma^i$) degrees of freedom [$\text{diag} (0, -1, 0, 1) = (1/2)(\sigma^z-\sigma^0)\otimes \Sigma^z$ and, $\text{diag} (-1, -1, 1, 1) =-\sigma^0\otimes \Sigma^z$] and then using its anticommuting ($\{ \sigma^i, \sigma^j\}=2\delta_{i,j}$ and $\{ \Sigma^i, \Sigma^j\}=2\delta_{i,j}$) properties in the calculation of orbital-current [See details in appendix \ref{AppendixB}]. It is important to notice that, the similarities between orbital-current operators of Eqs. (\ref{JorbAtomic}, \ref{JorbBerry}) are an artifact of the low-energy description of TMD. This similarity is not expected in calculations that include all Bloch bands of material \cite{Manchon-Orbital}. 

As done before, we use this operator together with the corrected energies [Eqs. (\ref{CorrectedEnergies}, \ref{correctionVal}, \ref{correctionCond})] and eigenstates [Eq. (\ref{CorrectedStates})] to compute the orbital-weighted Berry curvatures, $\Omega^{\text{Tot}}_{c(v),\pm}(\vec{q})$.  After performing the integral over azimuthal angle and doing the expansion in linear order on $t_{\perp}/\Delta$, the integrals are given by
\begin{eqnarray}
&&\int_0^{2\pi} \frac{d\theta}{(2\pi)} \Omega^{\text{Tot}}_{v,\pm}(\vec{q}) \approx -\frac{e}{\hbar \mu_B} \Bigg[ \frac{2a^4t^4\Delta^2}{(4a^2t^2q^2+\Delta^2)^{5/2}} \nonumber \\
&&  \pm t_{\perp} \frac{2a^4t^4\Delta^2\left(4a^2t^2q^2+\Delta^2+\Delta \sqrt{4a^2t^2q^2+\Delta^2} \right)}{\left(4a^2t^2q^2+\Delta^2 \right)^4} \Bigg] \nonumber \\ \label{BerryOmegaVal}
\end{eqnarray}
for the valence band and
\begin{eqnarray}
&&\int_0^{2\pi} \frac{d\theta}{(2\pi)} \Omega^{\text{Tot}}_{c,\pm}(\vec{q}) \approx -\frac{e}{\hbar \mu_B} \Bigg[ -\frac{2a^4t^4\Delta^2}{(4a^2t^2q^2+\Delta^2)^{5/2}} \nonumber \\
&& \pm t_{\perp} \frac{2a^4t^4\Delta^2\left(4a^2t^2q^2+\Delta^2-\Delta \sqrt{4a^2t^2q^2+\Delta^2} \right)}{\left(4a^2t^2q^2+\Delta^2 \right)^4} \Bigg] \nonumber \\ \label{BerryOmegaCond}
\end{eqnarray}
for the conduction band. Again, we evaluate the radial integration assuming $q\rightarrow \infty$ in the lower limit of integral and find the OHC, 
\begin{eqnarray}
\sigma^{\text{Tot}}_{OH}(E_f)=\sum_{\nu=\pm} \left[\sigma^{\text{Tot}}_{v,\nu} (E_f)+\sigma^{\text{Tot}}_{c,\nu} (E_f) \right]. \label{SigmaBerry}
\end{eqnarray}
\begin{widetext}
The contribution of valence band $\sigma^{\text{Tot}}_{v,\pm}$ is then
\begin{eqnarray}
\sigma^{\text{Tot}}_{v,\pm} &&(E_f)=g_s\left(\frac{e^2}{2\pi \hbar \mu_B}\right) \Bigg[\frac{a^2t^2\Delta^2}{6\left(4a^2t^2q^2_{v,\pm}+\Delta^2\right)^{3/2}} \pm t_{\perp} \frac{\Delta^2 a^2t^2 \left(4\Delta+5\sqrt{4a^2t^2q^2_{v,\pm}+\Delta^2} \right)}{40\left(4a^2t^2q^2_{v,\pm}+\Delta^2 \right)^{5/2}}\Bigg], \label{SigmaBerryVal}
\end{eqnarray} 
and the contribution of conduction band, given by $\sigma^{\text{Tot}}_{c,\pm} (E_f)$, is
\begin{eqnarray}
\sigma^{\text{Tot}}_{c,\pm} (E_f)=g_s\left(\frac{e^2}{2\pi \hbar\mu_B}\right) && \Bigg[-\frac{a^2t^2}{6\Delta}+\frac{a^2t^2\Delta^2}{6\left(4a^2t^2q^2_{c,\pm}+\Delta^2\right)^{3/2}} \nonumber \\ 
&&\mp t_{\perp} \frac{a^2t^2 \left(\Delta^5+\left(4a^4t^4q^4_{c,\pm}+2a^2t^2q^2_{c,\pm}\Delta^2 -\Delta^4\right)\sqrt{4a^2t^2q^2_{c,\pm}+\Delta^2} \right)}{10\Delta^2\left(4a^2t^2q^2_{c,\pm}+\Delta^2 \right)^{5/2}}\Bigg].\label{SigmaBerryCond}
\end{eqnarray}
The Fermi momenta $q_{v(c),\pm}$ are given by Eq. (\ref{FermiMomenta}). Again, we have included the spin-degeneracy $g_s=2$.
\end{widetext}

In Fig. \ref{fig:fig2}, we show the orbital Hall conductivities of Eqs. (\ref{SigmaBerry}, \ref{SigmaBerryVal}, \ref{SigmaBerryCond}) as a function of Fermi energy, using the parameters of the bilayer MoS$_2$. Similar to calculations using the intra-atomic approximation for the OAM operator here, we also obtain an OHC plateau when the Fermi energy lies inside the insulating gap. Nevertheless, in the case of an OAM operator described by orbital magnetic moments of Bloch states, the height of the OHC plateau is not quantized and assumes the value 
\begin{eqnarray}
\overline{\sigma}^{\text{Tot}}_{OH}=\frac{2a^2t^2}{3\Delta} \left(\frac{2e^2}{2\pi \hbar \mu_B}\right)=\frac{4}{3} \frac{\mu^*_B}{\mu_B}  \left( \frac{e}{2\pi}\right).
\label{SigmaBerryGap} 
\end{eqnarray}
As obtained in Ref. \cite{OHE_Bhowal-Vignale} for the gapped graphene monolayer, the height of OHC of 2H-TMDs within this method scales with $1/\Delta$. Generically speaking, we can understand this behavior of OHE with the inverse of bandgap as a reminiscent effect of the geometry of Bloch bands encoded in the orbital magnetic moment description of the OAM operator. It is easy to note that, for small bandgaps, $m_0(q)$ in the Eq. (\ref{m0}) is a representation of the Dirac-delta function, $\lim_{\epsilon \rightarrow 0}=\epsilon /(x^2+\epsilon^2)=\pi \delta(x)$. This diverging behavior propagates after $q$-integration up to the result of the Eq. (\ref{SigmaBerryGap}). It is worth mentioning that limit of vanishing bandgap makes sense only for graphene, which is a gapless material in a pristine situation. It is possible to induce controllable bandgaps in graphene by proximity to distinct materials \cite{graphene-Gap-hbn, Eu-Crystal-Field, Eu-QHE}. In the case of the bilayer of 2H-TMD, the bandgap is intrinsic to the band structure of the material [Fig. \ref{fig:fig1} (c)]. For this reason, $\Delta$ cannot be seen as a free parameter in the context of the present work, assuming a fixed value obtained by fitting density functional calculations. In the second equality of Eq. (\ref{SigmaBerryGap}), we used the atomic and renormalized Bohr magnetons defined previously to write the height of the OHC plateau in a more transparent expression. The atomic Bohr magneton ($\mu_B$) depends on fundamental constants while the renormalized Bohr magneton ($\mu^*_B$) depends on the specific band structure of a given material.  In the case of massive Dirac fermions, $\mu^*_B$ has the dependence with $1/\Delta$. Substituting in Eq. (\ref{SigmaBerryGap}) the parameters $a=3.160 \angstrom$, $t=1.137 \text{eV}$ and $\Delta=1.766 \text{eV}$ of the bilayer of 2H-MoS$_2$ presented in Sec. \ref{sec2}, we obtain the result $\overline{\sigma}^{\text{Tot}}_{OH} \approx 2.52  \left( e/ 2\pi \right)$.

\subsection{Discussion}

Here, we discuss some of the physical consequences of the results derived in the previous two subsections, giving special attention to Eqs. (\ref{SigmaAtomicGap}) and (\ref{SigmaBerryGap}). Despite the quantitative difference in height of the OHC plateaus, both formalisms for OAM discussed above predict the existence of finite OHE inside the insulating gap of unbiased bilayers of 2H-TMDs. This qualitative agreement between the OHC obtained via usual intra-atomic approximation [Eq. (\ref{SigmaAtomicGap})] and the OHC computed via orbital magnetic moment description of OAM [Eq. (\ref{SigmaBerryGap})] should be considered one of the main messages of this work. Previous works interpreted the transport of OAM in TMDs in terms of the valley Hall effect, assuming that it also generates transport of magnetic moment, once we can associate a magnetic moment with inverted signals to the inequivalent valleys \cite{VHE-graphene-XIAO}. This picture may explain the experimental results in noncentrosymmetric systems such as monolayer TMDs and gapped graphene. On the other hand, this interpretation led previous works to assume that centrosymmetric systems, such as unbiased bilayer TMDs, should not exhibit transport of OAM \cite{VHE-Momentum_2, VHE-Momentum_3, VHE-Momentum_4}. This assumption follows from the absence of orbital magnetic moment in Bloch-states of materials that preserve both time-reversal and spatial inversion symmetries [see the discussion in Appendix \ref{Appendix}]. Experimental works attributed the observed signal of OAM transport in unbiased bilayer TMDs to a break of spatial inversion symmetry induced by the substrate \cite{VHE-Momentum_2, VHE-Momentum_3, VHE-Momentum_4}. This argument was used even in situations where the material of substrate ($h$-BN) should not interact significantly with the bilayer of 2H-TMD \cite{VHE-Momentum_5}. In addition, the valley Hall effect has some inherent conceptual problems, such as the need of an artificial separation of the BZ \cite{OHE_Bhowal-Vignale}. The results of Eqs. (\ref{SigmaAtomicGap}) and (\ref{SigmaBerryGap}) above show that the transport OAM can occur in centrosymmetric bilayers of 2H-TMDs, as anticipated in Ref. \cite{Us3-PRL} by using intra-atomic approximation.

It is worth mentioning that we cannot understand the transport of OAM in the bilayer of 2H-TMD in terms of the valley Hall effect of individual layers \cite{Comment}. For any infinitesimally small but finite interlayer hopping $t_{\perp}$, the states of individual layers combine to form the bonding and antibonding superpositions , as seen in Eq. (\ref{CorrectedStates}) \cite{Reply}. These states have no intrinsic magnetic moment, a consequence of the inversion symmetry of bilayer 2H-TMD [see Fig. \ref{fig:fig1} (a)]. Nevertheless, the OHE is possible if one uses the non-Abelian structure of the orbital magnetic moment to describe the OAM operator, as we detail in appendix \ref{Appendix}, leading to the results presented in Sec. \ref{SecOHEBerry}. The non-Abelian structure of the magnetic moment operator emerges for nearly degenerate bands \cite{Bloch-Orbital-Moment-Culcer-3} [Fig. \ref{fig:fig1} (c)] and it is essential for the understanding of the OHE in unbiased bilayer TMDs.

\section{Effect of the potential-asymmetry between layers \label{sec4}}

To go further in the comparison between qualitative predictions of OHE using intra-atomic approximation and orbital magnetic moment description of OAM operator, we include a spatial-inversion symmetry breaking term in the Hamiltonian. This is achieved by including a term that mimics the effect of the gate voltage (bias) used in experiments. Non-perturbative numerical calculations in intra-atomic approximation show that the inclusion of a gate bias does not affect the height of the OHC plateau \cite{Us3-PRL}. Here, we use perturbation theory to test the effect of gate voltage in OHE using both descriptions of the OAM operator studied in this work. The calculations are analogous to those described in the previous section, so we only present the main results. 

\subsection{Perturbation theory for biased bilayer of 2H-TMD}
We include the gate voltage term in the unperturbed Hamiltonian,
\begin{eqnarray}
H^U_0(\vec{q})&=&\begin{bmatrix}
\Delta+U & \gamma_+  & 0 & 0\\
  \gamma_- & +U & 0 & 0\\
0 & 0 & \Delta-U & \gamma_-\\
0 & 0 & \gamma_+ & -U
\end{bmatrix}.\label{H0U}
\end{eqnarray}
The finite gate voltage ($U\neq 0$) breaks the inversion symmetry [Fig. \ref{fig:fig1} (a)] of the bilayer system, generating a bias between the individual layers. We still consider the interlayer hopping term [Eq. (\ref{H1})] as a perturbation. But, contrary to the case of unbiased bilayer of 2H-TMD where the interlayer hopping plays an essential role, for the case of the biased bilayer, the interlayer hopping does not generate corrections in the first order on $t_{\perp}$. Diagonalizing the Hamiltonian of Eq. (\ref{H0U}), we obtain the energy dispersions   
\begin{eqnarray}
\epsilon^{0,U}_{v,1(2)}(q)=\frac{1}{2}\left(\Delta \pm 2U - \sqrt{\Delta^2+4a^2t^2q^2} \right), \label{E0UVal} \\
\epsilon^{0,U}_{c,1(2)}(q)=\frac{1}{2}\left(\Delta \pm 2U + \sqrt{\Delta^2+4a^2t^2q^2} \right). \label{E0UCond}
\end{eqnarray}
These solutions correspond to a hard shift $+U$ ($-U$) in the energy of layer 1(2) of the unbiased bilayer [Eqs. (\ref{E0})]. The unperturbed eigenstates are not affected by the gate potential and we obtain the same eigenvectors of Eqs. (\ref{PsiVal01}, \ref{PsiVal02}, \ref{PsiCond01}, \ref{PsiCond02}), i.e., $\big|\psi^{U}_{1(2),\tau,v(c)}\rangle= \big|\psi_{1(2),\tau,v(c)}\rangle$. The energy spectrum of Eqs. (\ref{E0UVal}, \ref{E0UCond}) is non-degenerate for finite $U$. So, to include the effects of the interlayer hopping term [Eq. (\ref{H1})], we can apply the non-degenerate perturbation theory. Computing the correction on energy, we found that the interlayer hopping does not affect the spectra in first-order perturbation theory, $\langle \psi^{U}_{1(2),\tau,v(c)}\big| H_1\big| \psi^{U}_{1(2),\tau,v(c)}\rangle=0$. \subsection{OHE in biased bilayer TMD}

\begin{figure}[h]
	\centering
	 \includegraphics[width=0.9\linewidth,clip]{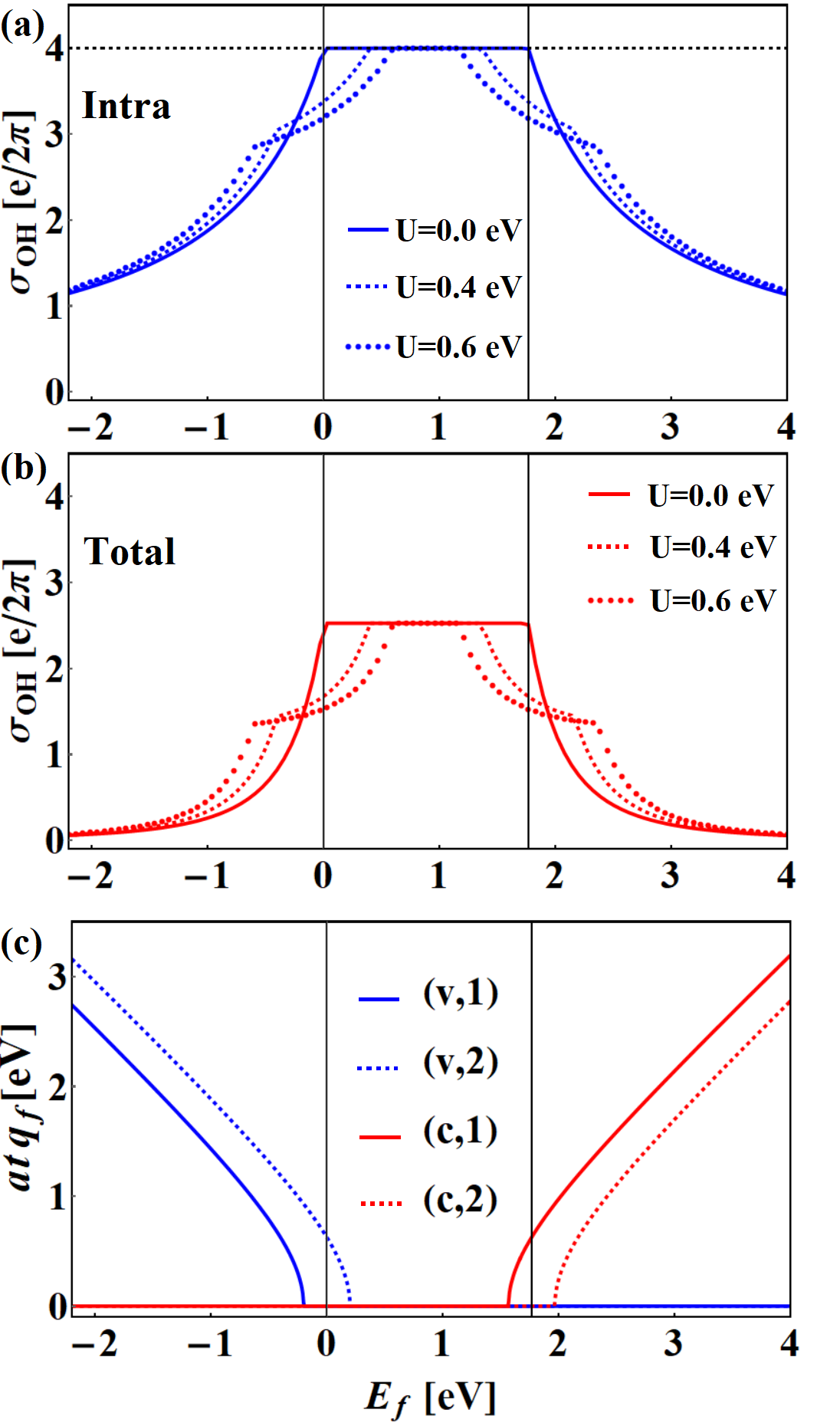}   
	\caption{The effect of inversion symmetry-breaking gate potential in OHC computed using intra-atomic approximation [Eqs. (\ref{SigmaAtomU}), (\ref{SigmaAtomValU}), and (\ref{SigmaAtomCondU})] (a) and Bloch state orbital magnetic moment description of OAM [Eqs. (\ref{SigmaBerryU}), (\ref{SigmaBerryValU}), and (\ref{SigmaBerryCondU})] (b). (c) Fermi-momentum [Eqs. (\ref{qUv}, \ref{qUc})], as a function of Fermi-energy, for the valence and conduction band states in bilayer of 2H-MoS$_2$ with an gate voltage bias $U=0.2 \text{eV}$. \new{In panel (c), $a=3.160 \ \angstrom$ and $t=1.137 \ \text{eV}$ [see Sec. \ref{sec2}].}}
	\label{fig:fig3}
\end{figure}

Following the same steps of Sec. \ref{sec3}, we can obtain analytical expressions for the OHC in terms of the Fermi momenta. The Fermi momenta for biased bilayer are given by
\begin{eqnarray}
q^U_{v,1(2)}= -\frac{\sqrt{E_f\pm U}\sqrt{E_f\pm U-\Delta}}{a t}, \ \ \left(\text{for} E_f\in \text{VB} \right),\nonumber \\
 \label{qUv} \\
q^U_{c,1(2)}= +\frac{\sqrt{E_f\pm U}\sqrt{E_f\pm U-\Delta}}{a t}. \ \ \left(\text{for} E_f\in \text{CB} \right), \nonumber \\ \label{qUc}
\end{eqnarray}
$E_f\in \text{VB(CB)}$ means that the equation holds for $E_f$ crossing the respective state of the valence (conduction) band and is zero otherwise. Fig. \ref{fig:fig3} (c) shows the Fermi momenta of Eqs. (\ref{qUv}, \ref{qUc}) as a function of the Fermi energy for a finite value of the gate voltage. The OAM operator in the intra-atomic approximation for biased bilayer of 2H-TMDs remains the same as the unbiased case [Eq. (\ref{LZAtomic})]. Repeating the steps of Sec. \ref{SecOHEAtomic}, we obtain the orbital Hall conductivity given by
\begin{eqnarray}
\sigma^{U,\text{Intra}}_{OH}(E_f)=\sum_{l=1,2} \left[\sigma^{U,\text{Intra}}_{v,l} (E_f)+\sigma^{U,\text{Intra}}_{c,l} (E_f)\right], \label{SigmaAtomU}
\end{eqnarray}
where,
\begin{eqnarray}
&&\sigma^{U, \text{Intra}}_{v,1(2)} (E_f)=g_s\frac{e}{2\pi}\left[ \frac{\Delta}{\sqrt{4\big(atq^U_{v,1(2)}\big)^2+\Delta^2}} \right], \label{SigmaAtomValU} \\
&&\sigma^{U,\text{Intra}}_{c,1(2)} (E_f)=-g_s\frac{e}{2\pi}\left[1-\frac{\Delta}{\sqrt{4\big(atq^U_{c,1(2)}\big)^2+\Delta^2}} \right]. \nonumber \\\label{SigmaAtomCondU}
\end{eqnarray}
In Fig. \ref{fig:fig3} (a), we show these orbital Hall conductivities as a function of the Fermi energy for different gate voltages. For biased bilayers, contrary to the unbiased case, we can apply the usual formula for nondegenerate bands [Eq. (\ref{MMNondegenerate})] described in the appendix \ref{Appsec1} to compute the orbital magnetic moment operator. We obtain the orbital magnetic moment on TB basis $\beta_{tb}$, $\hat{m}^U_{tb}(q)= \tau m_0(q)  \text{diag}\left(1, 1, -1, -1\right)$. The OAM operator reads, $L_z^{U,Tot}(q)= (-\hbar/\mu_B)  \hat{m}^U_{tb}(q)$, having the same form as the one obtained in the unbiased case. After following similar steps described in Sec. \ref{SecOHEBerry}, we obtain the OHC in the orbital magnetic moment description,
\begin{eqnarray}
\sigma^{U,\text{Tot}}_{OH}(E_f)=\sum_{l=1,2} \left[ \sigma^{U,\text{Tot}}_{v,l} (E_f)+\sigma^{U,\text{Tot}}_{c,l} (E_f) \right], \label{SigmaBerryU}
\end{eqnarray}
where,
\begin{eqnarray}
\sigma^{U,\text{Tot}}_{v,1(2)} &&(E_f)=g_s\left(\frac{e^2}{2\pi \hbar \mu_B}\right)  \nonumber \\
&&\left[\frac{a^2t^2\Delta^2}{6\left(4\big(atq^U_{v,1(2)}\big)^2+\Delta^2\right)^{3/2}}\right], \label{SigmaBerryValU}  \\
\sigma^{U,\text{Tot}}_{c,1(2)} &&(E_f)=g_s\left(\frac{e^2}{2\pi \hbar \mu_B}\right)  \nonumber \\ 
&&\left[-\frac{a^2t^2}{6\Delta}+\frac{a^2t^2\Delta^2}{6\left(4\big(atq^U_{c,1(2)}\big)^2+\Delta^2\right)^{3/2}}\right]. \label{SigmaBerryCondU}
\end{eqnarray}
Fig. \ref{fig:fig3} (b) shows the orbital Hall conductivities with the orbital magnetic moment description for three values of gate voltage. Comparing the plots of panels (a) and (b) of Fig. \ref{fig:fig3}, we conclude that the effects of a finite gate bias on OHC predicted by intra-atomic approximation and by Bloch state orbital magnetic moment description of OAM qualitatively agree. Particularly the height of the OHC plateau is not affected by the intensity of the gate potential \cite{Us3-PRL}. 

Generically speaking, the spatial inversion asymmetry caused by gate bias can induce an orbital-Rashba coupling in bilayer systems \cite{Orbital-Rashba, Importanceof-ModTheo-4}. This effect can lead to the appearance of orbital textures that can be observed with photoemission spectroscopy techniques \cite{OR-ARPES-theory-1, OR-ARPES-theory-2, Orbital-Texture-Exp-TMD-1, Orbital-Texture-Exp-TMD-2, Orbital-Texture-Exp-TMD-3} and may affect the transport of OAM \cite{Us1-PRB, Us2-PRBR}. The intensity of the orbital-Rashba effect depends on the inter-orbital hybridization between nearest neighbors atoms. In the supplementary material of Ref. \cite{Us3-PRL}, the effect of gate bias in the intra-atomic approximation is considered using a tight-binding model that includes the complete orbital structure of chalcogen and the transition metal of the MoS$_2$. The results obtained using this model agrees with the one obtained using the Mo (d)-orbital low-energy model used here, suggesting that, for MoS$_2$, the orbital-Rashba effect induced by gate voltage is weak. For other compounds of TMD family, the orbital-Rashba effect may be relevant and modifies the height of the orbital Hall plateau. Unless the energy scale of the orbital-Rashba effect reaches the energy scale of the insulating gap of the compound, we would not expect the vanishing of the orbital Hall plateau discussed here \cite{Kane-Mele-SHE}. This point is beyond the scope of the present paper. Nonetheless, this is an interesting possibility to be explored in future works.

\section{Final remarks and conclusions \label{sec5}}

To summarize, we presented a detailed study of the orbital Hall effect in bilayer TMDs with 2H stacking. We took into account the effect of interlayer hopping $t_{\perp}$ using first-order perturbation theory. This allowed us to obtain analytical expressions for the orbital Hall conductivity as a function of Fermi energy, corrected up to linear order in $t_{\perp}/\Delta$. We then compared the results of  the orbital Hall conductivity given by two different descriptions of the orbital angular momentum operator. The first one is the well known intra-atomic approximation that takes into account only the intrasite contribution of the wave function. The second approach describes the orbital angular momentum operator in terms of the intrinsic orbital magnetic moment of the Bloch states. This approach includes the contributions coming from intersite and intrasite parts of the electronic wave function \cite{Murakami-IntersiteIntrasite, ModTheo-5}. We found that, despite a natural quantitative discrepancy, both methods agree on their qualitative predictions. Particularly, they agree in predicting a plateau of orbital Hall conductivity inside the insulating gap of the bilayer TMD that is robust against a gate voltage that breaks inversion symmetry. As we detailed in the main text, taking into account the interlayer hopping $t_{\perp}$ on the bilayer Hamiltonian is essential to understand the orbital Hall effect in the unbiased bilayer TMD. If one makes $t_{\perp}=0$ in Eq. (\ref{Heff}), the Hilbert spaces of the individual layers become decoupled and the orbital Hall conductivity of the bilayer is exactly twice of monolayers. When $t_{\perp}\neq 0$, no matter how small it is, the wave functions of bilayer change dramatically, becoming superpositions of wave functions in the two layers. The form of the orbital magnetic moment also changes dramatically, going from diagonal to the off-diagonal matrix elements in the non-Abelian description [see Appendix \ref{Appsec2}]. Numerically, the corrections of terms proportional to $t_{\perp}$ in the final expressions of orbital Hall conductivity [Eqs. (\ref{SigmaAtom},\ref{SigmaAtomVal}, \ref{SigmaAtomCond}) and Eqs. (\ref{SigmaBerry}, \ref{SigmaBerryVal}, \ref{SigmaBerryCond})] are indeed small. The importance of interlayer hopping in the description of the orbital Hall effect in bilayers of 2H-TMDs is from the conceptual perspective. Once it is finite and cannot be turned off in the bilayer, the Hilbert space of the layers becomes connected (see Ref. \cite{VHE-Momentum_2}). In this situation, due to the spatial-inversion symmetry of the bilayer, the non-Abelian nature of the orbital magnetic moment operator turns out to be essential to the appearance of finite orbital Hall conductivity. If one does not consider the non-Abelian structure of the orbital magnetic moment operator, the orbital Hall effect [for $t_{\perp}\neq 0$ in Eq. (\ref{Heff})] within this approach is enforced to vanish by symmetry constraints in its diagonal elements [Eq. (\ref{MMDegenerateMatrix})]. The non-Abelian description of orbital angular momentum in bilayers of 2H-TMDs was neglected in previous experimental works that assumed the absence of orbital angular moment transport in unbiased system \cite{VHE-Momentum_2, VHE-Momentum_3, VHE-Momentum_4, VHE-Momentum_5}. In intra-atomic approximation, this issue does not occur and a finite orbital Hall effect naturally appears, as shown in Ref. \cite{Us3-PRL}.

The orbital magnetic moment of Bloch states we used to define the orbital angular momentum current can be related with self-rotation of electronic wave-packet or Wannier function \cite{ModTheo-2, ModTheo-3, ModTheo-5}. The accumulation of this self-rotation contribution generated by the orbital Hall effect could, in principle, be measured by magnetic circular dichroism \cite{ModTheo-5}. Once the approach based on Bloch states orbital magnetic moment contains contributions coming from both intrasite and intersite parts of wave functions, it should be directly compared with intra-atomic approximation. It is well known that the accuracy of the intra-atomic approximation is highly dependent on the specificities of the material \cite{Importanceof-ModTheo-1, Importanceof-ModTheo-2, Importanceof-ModTheo-3}. In our work, we use the low-energy model for bilayers of 2H-TMDs elaborated in Refs. \cite{Low-Energy-bilayer, 3-Bands-Model}. These references construct this low-energy Hamiltonian from a tight-binding model with parameters adjusted to fit density functional calculations. One can construct this low-energy Hamiltonian of TMDs from calculations using maximally localized Wannier functions which present an accurate description of the orbital angular momentum of material \cite{Low-energy_x_Wannier-TMD}. With this construction, the information on orbital angular momentum is encoded in the band structure and the wave function and reflected in the low-energy Hamiltonian when expanded near valleys of the Brillouin zone. Without this, the compatibility between methods is not guaranteed.

The key results of our work are the Eqs. (\ref{SigmaAtomicGap}) and (\ref{SigmaBerryGap}) and the insulating orbital Hall conductivity plateaus of the Figs. \ref{fig:fig2} and \ref{fig:fig3}. The analytical expressions derived here for orbital conductivity in metallic regimes (Fermi energy crossing electronic bands) may help to understand some experimental results in clean samples but are, in general, subject to the effects of the disorder \cite{Vertex-Dimitrova, Vertex-Ebert, Vertex-Raimondi}. On the other hand, the effect of disorder should not affect the transport of orbital angular momentum when the Fermi energy lies inside the insulating gap \cite{Us1-PRB}. Finally, we mention that the theoretical analysis exposed in this work suggests the orbital Hall effect as a more appropriate description of the transport of orbital angular momentum in bilayers TMDs \cite{Us3-PRL, OHE_Bhowal-Vignale}, in contrast to previous literature based on the valley Hall effect. Our results also show that the orbital Hall effect cannot be seen as a hidden valley Hall effect, as done in Ref. \cite{Comment}, due to the finite interlayer hopping in the bilayer system \cite{Reply}. Our results may also give new insights on previous experimental results on valley Hall effect on TMDs and motivate new experiments focused on the characterization of the orbital Hall effect in these systems.

\begin{acknowledgments}
	We acknowledge CNPq/Brazil, CAPES/Brazil, FAPERJ/Brazil and INCT Nanocarbono for financial support. TGR acknowledges funding from Fundação para a Ciência e a Tecnologia and Instituto de Telecomunicações - grant number UID/50008/2020 in the framework of the project Sym-Break. SB thanks ETH, Zurich for financial support. 
\end{acknowledgments}	


\appendix
\section{The orbital magnetic moment of Bloch-states \label{Appendix}}
\subsection{General theory \label{Appsec1}}
As we mentioned in the main text, Bloch electrons may carry an intrinsic orbital magnetic moment. It is also known as valley magnetic moment in the literature of 2D materials. Here, we review the theory of intrinsic orbital magnetic moment of Bloch bands developed on Refs. \cite{Bloch-Orbital-Moment-Khon-1, Bloch-Orbital-Moment-Chang-2, Bloch-Orbital-Moment-Culcer-3, Bloch-Orbital-Moment-Review-1, Bloch-Orbital-Moment-Review-2}. More recently, this theory found connection with the so-called modern theory of orbital magnetization based on the framework of the Berry phase formalism \cite{ModTheo-1, ModTheo-2, ModTheo-3, ModTheo-4, vanderbiltBook}. Here we focus on 2D systems but, the extension to higher dimensions follows straightforwardly. We treat the cases of nondegenerate \cite{Bloch-Orbital-Moment-Khon-1, Bloch-Orbital-Moment-Chang-2} and nearly-degenerate \cite{Bloch-Orbital-Moment-Culcer-3, Bloch-Orbital-Moment-Review-1, Bloch-Orbital-Moment-Review-2} Bloch bands separately. 

\subsubsection{Nondegenerate band}
First, we consider the case of a nondegenerate band $n$ with dispersion relation $E_{n,\vec{k}}$ and, periodic part of Bloch function $\big|u_{n,\vec{k}}\rangle$. The intrinsic orbital magnetic moment of this Bloch band is given by,
\begin{eqnarray}
m^z(\vec{k})=-i\frac{e}{2\hbar}\langle \nabla_{\vec{k}} u_{n,\vec{k}}\big| \times \left(\hat{\mathcal{H}}(\vec{k})-E_{n,\vec{k}} \hat{\mathbb{1}}\right) \big|\nabla_{\vec{k}} u_{n,\vec{k}} \rangle, \nonumber \\ \label{MMNondegenerate}
\end{eqnarray}
where, $\nabla_{\vec{k}}=\partial_{k_x}\hat{x}+\partial_{k_y}\hat{y}$. This orbital magnetic moment transforms under time-reversal symmetry like, $\mathcal{T}: m^z(\vec{k})\rightarrow -m^z(-\vec{k})$. The transformation over spatial-inversion symmetry is, $\mathcal{I}: m^z(\vec{k})\rightarrow +m^z(-\vec{k})$. Therefore, in systems that preserve, simultaneously, the time-reversal and the spatial inversion symmetries, Eq. (\ref{MMNondegenerate}) gives $m^z(\vec{k}) \rightarrow 0$. To obtain a finite orbital magnetic moment from Eq. (\ref{MMNondegenerate}) is necessary to break spatial inversion or time-reversal symmetry. This occurs in monolayers of H-TMDs in which the spatial-inversion symmetry is absent, and Eq. (\ref{MMNondegenerate}) result in $\tau m_0(q)$, where $m_0(q)$ was defined in sec. \ref{SecOHEBerry}. In the case of the bilayer of 2H-TMD, the spatial-inversion symmetry is restored [see Fig. \ref{fig:fig1} (a)] and, Eq. (\ref{MMNondegenerate}) gives $m^z(\vec{k}) \rightarrow 0$ for any finite value of interlayer hopping $t_{\perp}$. In this case, due to the nearly degenerate band structure of the bilayer, the use of the matrix form of orbital magnetic moment operator discussed below become necessary.

\subsubsection{Nearly degenerate bands}
Now, we follow Refs. \cite{Bloch-Orbital-Moment-Culcer-3, Bloch-Orbital-Moment-Review-1, Bloch-Orbital-Moment-Review-2} and consider the case of a set of $N$ Bloch states $\big|u_{n,\vec{k}}\rangle$ with very close energies $E_{n,\vec{k}}$, where $n=1,2,...,N$. The matrix-element of the magnetic-moment operator between two states $\big|u_{n,\vec{k}}\rangle$ and $\big|u_{m,\vec{k}}\rangle$ in this nearly-degenerate subspace is,
\begin{eqnarray}
m&&_{n,m}^z(\vec{k})=\nonumber \\ 
&&-i\frac{e}{2\hbar}\langle \nabla_{\vec{k}} u_{n,\vec{k}}\big| \times \left[ \hat{\mathcal{H}}(\vec{k})-\left(\frac{E_{n,\vec{k}}+E_{m,\vec{k}}}{2}  \right)\hat{\mathbb{1}} \right] \big|\nabla_{\vec{k}} u_{m,\vec{k}} \rangle \nonumber \\
\label{MMDegenerate}
\end{eqnarray}
The diagonal elements ($m=n$) reduce to Eq. (\ref{MMNondegenerate}), obeying the same constraints imposed by spatial inversion and time-reversal symmetries. For nearly degenerate bands, the orbital magnetic moment acquires a non-Abelian structure \cite{Bloch-Orbital-Moment-Culcer-3}, being described by a $N$-dimensional matrix, 
\begin{eqnarray}
\hat{m}^z(\vec{k})= \begin{bmatrix}
m_{1,1}^z(\vec{k}) & m_{1,2}^z(\vec{k}) & \ldots & m_{1,N}^z(\vec{k})\\
m_{2,1}^z(\vec{k})  & m_{2,2}^z(\vec{k}) & \ldots & m_{2,N}^z(\vec{k}) \\
\vdots & \vdots & \ddots  & \vdots \\
m_{N,1}^z(\vec{k}) & m_{N,2}^z(\vec{k}) & \ldots & m_{N,N}^z(\vec{k}) 
\end{bmatrix}, \label{MMDegenerateMatrix}
\end{eqnarray}
where the matrix is written on the basis of Bloch eigenstates $\beta_u=\{u_1; u_2; ... ; u_N \}$. In contrast to the diagonal matrix elements, the non-diagonal elements of Eq. (\ref{MMDegenerateMatrix}) are not constrained by symmetries discussed earlier. They can be nonzero even in systems that preserve both spatial-inversion and time-reversal symmetry. The condition of validity of the formalism used to derive Eq. (\ref{MMDegenerateMatrix}) is the absence of transitions, induced by perturbing fields, out of the $N$-dimensional manifold under consideration \cite{Bloch-Orbital-Moment-Culcer-3}. This translates into the necessity for the nearly-degenerate manifold $\big|u_{1..N,\vec{k}}\rangle$ to be energetically isolated from other bands. In what follows, we apply this formalism to the specific case of bilayer of 2H-TMD. The huge bandgap of the bilayer [see Fig. \ref{fig:fig1}] forbids the transitions between conduction and valence bands. This allows the application of the Eq. (\ref{MMDegenerateMatrix}) separately to the valence and conduction subspaces. 

\subsection{Application to unbiased bilayer of 2H-TMD \label{Appsec2}}
We now apply the theory of orbital magnetic moment of Bloch states for nearly degenerate bands [Eqs. (\ref{MMDegenerate}, \ref{MMDegenerateMatrix})] to the unbiased bilayer of 2H-TMD discussed in Sec. \ref{SecOHEBerry} of the main text. For this purpose, we use the states of Eq.(\ref{CorrectedStates}) and corrected energies of Eqs.(\ref{CorrectedEnergies}, \ref{correctionVal}, \ref{correctionCond}) given by perturbation theory and the Hamiltonian of Eq. (\ref{Heff}) with $\lambda \rightarrow 0$. Due to the large band-gap separating the valence and conduction bands in the unbiased bilayers of 2H-TMDs, we can apply Eqs. (\ref{MMDegenerate}, \ref{MMDegenerateMatrix}) for these subspaces independently. on the basis $\beta_{\Phi}=\{ \Phi_{-,v}; \Phi_{+,v}; \Phi_{-,c}; \Phi_{+,c} \}$ of states given by perturbation theory [Eq. (\ref{CorrectedStates})], the magnetic moment matrix assumes a block-diagonal form,
\begin{eqnarray}
\hat{m}_{\Phi}^z(q)= \begin{bmatrix}
\hat{\mathbb{m}}_{v}^z(q) & \hat{\mathbb{0}}_{2\times 2} \\
\hat{\mathbb{0}}_{2\times 2} & \hat{\mathbb{m}}_{c}^z(q)
\end{bmatrix}, \label{MM2HTMD}
\end{eqnarray}
where, $\hat{\mathbb{0}}_{2\times 2}$ is a $2\times 2$ null matrix, and $\hat{\mathbb{m}}_{v}^z(q)$, $\hat{\mathbb{m}}_{c}^z(q)$ are the matrices computed in the subspace of valence and conduction states, respectively. Applying Eq. (\ref{MMDegenerate}), we obtain,
\begin{eqnarray}
\hat{\mathbb{m}}_{v}^z(q)= \hat{\mathbb{m}}_{c}^z(q)= \tau\begin{bmatrix}
0 & m_0(q) \\
m_0(q) & 0
\end{bmatrix}, \label{MM2HTMDValCond}
\end{eqnarray}
where $m_0(q)$ is given by Eq. (\ref{m0}). Contrary to the monolayer case, the bilayer of 2H-TMD possesses an inversion symmetry point $\mathcal{I}$ in interlayer space [See Fig. \ref{fig:fig1} (a)]. It also preserves the time-reversal symmetry $\mathcal{T}$. Therefore, for any small amount of interlayer hopping $t_{\perp}$, the states of individual layers combine in the bonding and antibonding superpositions of Eq. (\ref{CorrectedStates}), making the diagonal elements of the magnetic-moment matrix go to zero. It is possible to use a pictorial view, making use of projection operators, to interpret the vanishing of diagonal elements of the matrix of Eqs. (\ref{MM2HTMD}, \ref{MM2HTMDValCond}) in terms of the magnetic moment of monolayers with inverted signs. But, we cannot attribute a finite magnetic moment to eigenstates of the unbiased bilayer of 2H-TMD. The zero diagonal matrix elements of Eqs. (\ref{MM2HTMD}, \ref{MM2HTMDValCond}) made previous works conclude that no magnetic moment flows in unbiased bilayers of 2H-TMDs \cite{VHE-Momentum_2, VHE-Momentum_3, VHE-Momentum_4, VHE-Momentum_5}. As is discussed in the main text, one of the conclusions of this work is that the flow of magnetic moments is possible \cite{Us3-PRL} and, in this formalism, has its origins in off-diagonal matrix elements in Eqs. (\ref{MM2HTMD}, \ref{MM2HTMDValCond}). In addition, as anticipated in Ref. \cite{Reply}, the transport of magnetic moment in unbiased bilayer 2H-TMD cannot be understood in terms of the individual layers \cite{Comment}, being necessary to employ the scheme based on OHE introduced in Ref. \cite{OHE_Bhowal-Vignale} together with the non-Abelian structure of the magnetic moment matrix [Eqs. (\ref{MMDegenerateMatrix}, \ref{MM2HTMD}, \ref{MM2HTMDValCond})]. Finally, to use this operator in the formula of Eqs. (\ref{SigmaLR}, \ref{OBC}) for OHC, it is necessary to perform a unitary transformation to change the magnetic moment matrix from basis $\beta_{\Phi}$ [Eq. (\ref{MM2HTMD})] to tight-binding basis $\beta_{tb}$. After this, we obtain equation (\ref{mtb}) of the main text.
Introducing a constant to convert the magnetic-moment to the unity of OAM, we obtain the OAM operator in the orbital magnetic moment description Eq. (\ref{LZBerry}). To close this appendix, we make a brief technical comment. If one insists on calculating the matrix elements that connect valence and conduction bands, substituting them on off-diagonal blocks $\hat{\mathbb{0}}_{2\times 2}$ of the matrix of Eq. (\ref{MM2HTMD}), the results for OHC [Eqs. (\ref{SigmaBerry}, \ref{SigmaBerryVal}, \ref{SigmaBerryCond})] would not change. All the terms that emerge from these contributions vanish after integral over azimuthal angle [Eqs.(\ref{BerryOmegaVal}, \ref{BerryOmegaCond})]. Nevertheless, we reinforce that the non-Abelian structure of the magnetic moment should consider only states inside the nearly degenerate subspace \cite{Bloch-Orbital-Moment-Culcer-3}.

\section{Details on the calculation of orbital currents \label{AppendixB}}

As we mentioned in the main text, one striking feature of our work is the similarity of the matrix form of orbital currents [Eqs. (\ref{JorbAtomic}, \ref{JorbBerry})] within the two approaches, despite the difference in its OAM operators [Eqs. (\ref{LZAtomic}, \ref{LZBerry})]. To clarify this point, we derive here the orbital current operators of Eqs. (\ref{JorbAtomic}, \ref{JorbBerry}). To this end, we define the Pauli-matrices $\sigma^{x,y,z}$ associated with the $d$-orbital character of the wave function and the corresponding identity matrix $\sigma^0$. We also define the Pauli-matrices $\Sigma^{x,y,z}$ related to the layer degree of freedom and its corresponding identity $\Sigma^0$. With this, we write the OAM operators as
\begin{eqnarray}
\hat{L}_z^{\text{Intra}}=2 \hbar \tau \left(\frac{\sigma^z-\sigma^0}{2}\right)\otimes \Sigma^z, \label{PauliLZINTRA}
\end{eqnarray}
and  
\begin{eqnarray}
\hat{L}_z^{\text{Tot}}=\left(\frac{m_0(q)}{\mu_B}\right) \hbar \tau \left(-\sigma^0\right)\otimes \Sigma^z. \label{PauliLZTOT}
\end{eqnarray}
We obtain the velocity operator in the y-direction directly from the Hamiltonian of Eq. (\ref{Heff}): $\hat{v}_y=-(at)/(\hbar)\sigma^y\otimes \Sigma^z$. The Pauli-matrices have anticommuting properties $\{\sigma^i, \sigma^j\}=2\delta^{i,j}$, and $\{\Sigma^i,\Sigma^j\}=2\delta^{i,j}$, where $\delta^{i,j}$ is the Kronecker delta. The orbital current in Bloch state orbital magnetic moment approach is 
\begin{eqnarray}
J_y^{z,\text{Tot}}&=&\frac{at}{2}\left(\frac{m_0(q)}{\mu_B}\right)\Bigg[\begin{matrix} L_z^{\text{Tot}}\propto \\ \overbrace{ \tau\left(-\sigma^0\right)\otimes\Sigma^z } \end{matrix} \ \ \begin{matrix} v_y \propto \\ \overbrace{. (-\sigma^y)\otimes\Sigma^z} \end{matrix} \nonumber \\
&& \ \ \ \ \begin{matrix} \\ \ + \ \end{matrix} \begin{matrix} v_y \propto \\ \overbrace{(-\sigma^y)\otimes\Sigma^z} \end{matrix}\begin{matrix} L_z^{\text{Tot}}\propto \\ \overbrace{ .\tau\left(-\sigma^0\right)\otimes\Sigma^z } \end{matrix} \Bigg] \nonumber \\
&=& at \left(\frac{m_0(q)}{\mu_B}\right) \tau \Sigma^0\otimes \sigma^y, \label{JTotPauli}
\end{eqnarray}
and, in the intra-atomic approximation
\begin{eqnarray}
J_y^{z,\text{Intra}}&=& at\Bigg[\begin{matrix} L_z^{\text{Intra}}\propto \\ \overbrace{\tau\left(\sigma^z-\sigma^0\right)/2\otimes\Sigma^z } \end{matrix} \ \ \begin{matrix} v_y \propto \\ \overbrace{. (-\sigma^y)\otimes\Sigma^z} \end{matrix} \nonumber \\
&& \begin{matrix} \\ \ + \ \end{matrix} \begin{matrix} v_y \propto \\ \overbrace{(-\sigma^y)\otimes\Sigma^z} \end{matrix}\begin{matrix} L_z^{\text{Intra}}\propto \\ \overbrace{ .\tau\left(\sigma^z-\sigma^0\right)/2\otimes\Sigma^z } \end{matrix} \Bigg] \nonumber \\
&=& at \tau \Sigma^0\otimes \left[ -\sigma^z\sigma^y/2-\sigma^y\sigma^z/2+\sigma^y\right] \nonumber \\
&=&at \tau \Sigma^0\otimes \sigma^y, \label{JIntraPauli}
\end{eqnarray} 
where, in the second line, we used $\{\sigma^z, \sigma^y\}=0$. In matrix representation, Eq. (\ref{JTotPauli}) correspond to Eq. (\ref{JorbBerry}) and, Eq. (\ref{JIntraPauli}) correspond to Eq. (\ref{JorbAtomic}). The matrix structure of orbital currents within the two approaches is proportional to $\tau \Sigma^0 \otimes \sigma^y$. The difference between the two expressions for the OAM [Eqs. (\ref{PauliLZINTRA}, \ref{PauliLZTOT})] is proportional to $\Sigma^z\otimes \sigma^z$, which does not contribute to the orbital current operator because it anticommutes with the $y$-component of velocity.

\bibliographystyle{apsrev}
%

\end{document}